\documentclass[nofootinbib, superscriptaddress, amsmath,amssymb, aps, twocolumn]{revtex4-2}

\usepackage{graphicx}
\usepackage[normalem]{ulem}
\usepackage{dcolumn}
\usepackage{bm}
\usepackage{wasysym}
\usepackage{hyperref}
\usepackage[mathlines]{lineno}
\usepackage{xcolor}
\usepackage{dsfont}

\usepackage{gensymb}
\usepackage{enumitem}
\usepackage{multirow}
\usepackage{booktabs}
\usepackage{array}
\newcolumntype{P}[1]{>{\centering\arraybackslash}p{#1}}
\usepackage{amssymb}
\usepackage{adjustbox}
\usepackage{orcidlink}
\usepackage{float}
\usepackage{graphicx}
\usepackage{dcolumn}
\usepackage[caption=false]{subfig}

\usepackage{tikz,fp}
\usetikzlibrary{external,fixedpointarithmetic,decorations.pathmorphing,positioning,calc,fadings,decorations.markings,decorations.pathreplacing,patterns,arrows}

\newcommand{\aA}{\alpha_1}
\newcommand{\aB}{\alpha_2}
\newcommand{\bA}{\beta_1}
\newcommand{\bB}{\beta_2}
\newcommand{\dA}{\delta_1}
\newcommand{\dB}{\delta_2}
\newcommand{\arcosh}{\mathrm{arcosh}}
\tikzset{
  label distance=-3pt,
  >=stealth,
  inner sep=1.5pt,
  boson/.style={
    decoration={snake, segment length=2mm, amplitude=0.5mm},
    decorate,
  },
  scalar/.style={
    dashed
  }
}

\newcommand*{\defeq}{\mathrel{\vcenter{\baselineskip0.5ex \lineskiplimit0pt
                     \hbox{\scriptsize.}\hbox{\scriptsize.}}}
                     =}

\begin{document}
\title{Black-hole Scattering in Einstein–scalar–Gauss–Bonnet: Numerical Relativity Meets Analytics}
\author{Shaun Swain}
 \email{sxs2159@student.bham.ac.uk}
\affiliation{School of Physics and Astronomy and Institute for Gravitational Wave Astronomy,
University of Birmingham, Edgbaston, Birmingham, B15 2TT, United Kingdom}

\author{Tamanna Jain}%
 \email{tj317@cam.ac.uk}
\affiliation{%
LPENS, Département de physique, École normale supérieure, Université PSL, Sorbonne Université, Université Paris Cité, CNRS, 75005 Paris
}
\affiliation{%
 Department of Applied Mathematics and Theoretical Physics,University of Cambridge, Wilberforce Road CB3 0WA Cambridge, United Kingdom
}%

\author{Llibert Arest\'e Sal\'o}
 \email{llibert.arestesalo@kuleuven.be}
\affiliation{Instituut voor Theoretische Fysica, KU Leuven. Celestijnenlaan 200D, B-3001 Leuven, Belgium}
\affiliation{Leuven Gravity Institute, KU Leuven. Celestijnenlaan 200D, B-3001 Leuven, Belgium}


\begin{abstract}
    The study of hyperbolic binary black hole encounters yields an effective probe of the strong field regime of black holes, thus providing an additional channel to test General Relativity. 
    We study the scattering of two black holes in Einstein-scalar-Gauss-Bonnet gravity, a well-motivated effective field theory of gravity, by comparing the scattering angle obtained from the first fully nonlinear black hole scattering simulations with its effective-one-body analytic description. We obtain excellent agreement between analytics and numerics, exhibiting accurate capturing of strong-field scalar-gravitational dynamics. Our work paves the way towards semi-analytical waveform templates of compact object binaries in modified theories of gravity. 
\end{abstract}

\maketitle

{\textit{\textbf{Introduction}}---}
The direct detection of gravitational wave (GW) signals, with the first detection 
of a  black-hole (BH) binary
GW150914 by the LIGO-Virgo collaboration in 2015~\cite{LIGOScientific:2016aoc}, 
has established GW astronomy
as a new pathway for exploring the physics of our universe in the strong-gravity regime.
The subsequent multi-messenger observation of
the neutron-star (NS) binary system GW170817~\cite{LIGOScientific:2017vwq}
cemented a strong relation to short gamma ray bursts and efficient
production of heavy elements. While the new tests of general relativity (GR) using GWs~\cite{LIGOScientific:2020tif,LIGOScientific:2021sio,LIGOScientific:2025rid,LIGOScientific:2026fcf,LIGOScientific:2026qni,LIGOScientific:2026wpt} 
continue to add more evidence supporting the exceptional accuracy 
of Einstein's theory, both the observational and theoretical inconsistencies motivate 
the search for deviations from general relativity in extreme gravitational regimes.

One common approach in modifying GR is to add
all possible terms allowed by symmetry (after field redefinitions) to the leading order GR Lagrangian. The higher-order terms are organized in a derivative expansion and multiplied by dimensionful coupling constants. To ensure consistency, one must adopt 
an effective field theory (EFT) point of view, in the sense
that the truncated theory will only be meaningful as long
as we are in a regime in which we can neglect the higher
order terms in the derivative expansion~\cite{Burgess:2020tbq}.

Einstein-scalar-Gauss-Bonnet~(EsGB) gravity, in which a scalar field couples to the Gauss-Bonnet curvature invariant, has recently attracted a lot of attention. From an EFT perspective, it can be seen as the leading order, parity-invariant correction to GR when including an extra scalar degree of freedom. This theory is also known to endow BHs with scalar hair~\cite{Kanti:1995vq,Yunes:2011we,Sotiriou:2013qea,Sotiriou:2014pfa,Sotiriou:2015pka}, enhancing deviations 
from GR and leading to nonlinear dynamical phenomena, such as scalarization of compact objects~\cite{Damour:1993hw, Silva:2017uqg,Doneva:2017bvd,Dima:2020yac,Doneva:2022ewd}. 

It has only very recently been possible to evolve fully nonlinear numerical simulations in these higher derivative theories of gravity. In particular, the breakthrough by Kov\'acs and Reall~\cite{Kovacs:2020pns,Kovacs:2020ywu} enabled the construction of a well-posed initial value formulation of all weakly coupled Horndeski and Lovelock theories of gravity.  This led to the first evolutions of black hole spacetimes in EsGB, both in generalized harmonic coordinates~\cite{East:2020hgw,East:2021bqk} and in the moving-punctures approach~\cite{AresteSalo:2022hua,AresteSalo:2023mmd}, followed by others~\cite{Corman:2022xqg,Doneva:2023oww,Doneva:2024ntw,Corman:2024vlk,AresteSalo:2025sxc,Shum:2025lgp}, which have made it possible to find new GW signatures~\cite{Lara:2025kzj,Corman:2025wun,Capuano:2026lhs}.

In parallel, the two-body problem in massless scalar-tensor and EsGB theories of gravity has been extensively studied analytically using the post-Newtonian (PN)~\cite{Blanchet:2013haa} and the effective-one-body (EOB) formalism~\cite{Buonanno:1998gg},
primarily for bound orbits. The growing interest in applying EFT techniques to the classical two-body system has led to the study of unbound configurations (hyperbolic encounters), for which the scattering angle $\chi$ provides a gauge-invariant observable. The results of the EOB Hamiltonian up to 3PN order~\cite{Jain:2022nxs,Jain:2023fvt,Julie:2022qux} enabled the first investigations of hyperbolic encounters (and $\chi$) in these theories within the PN framework~\cite{Jain:2023vlf}. The scattering angle results using EFT techniques have recently been extended to third post-Minkowskian (3PM) order~\cite{Bernard:2026old, Almeida:2026abr}.

In this work, we compute the effective potential $w$ of two BHs up to 3PM order within the EOB formalism in EsGB\footnote{At this order, EsGB reduces to a massless scalar-tensor theory; however, black holes do not carry scalar charge in the latter.} using the 3PM results of~\cite{Bernard:2026old, Almeida:2026abr}. 
Furthermore, we perform {the first} fully nonlinear numerical relativity (NR) simulations of hyperbolic encounters. We then compare the scattering angle $\chi$
determined from NR to the analytical estimates from the EOB potential $w$ to determine
the accuracy and validity of our analytical results. 
Such EOB models, validated and informed by NR simulations, have been extensively developed for unbound systems in GR~\cite{Rettegno:2023ghr, Buonanno:2024vkx, Swain:2024ngs, Clark:2025kvu, Long:2025nmj} and can be used to construct semi-analytic waveform models for bound orbits \cite{Buonanno:2024byg, Damour:2025uka}, which are essential to generate banks of precise waveform templates for detection and parameter estimation of GW events.

We follow the conventions in Wald's book~\cite{Wald:1984rg}.
We set $G=c=1$, unless specified explicitly. 

\vskip 5pt
{\textit{\textbf{Theory}}---}
We consider shift-symmetric EsGB gravity with action
\begin{eqnarray}
    S=\frac{1}{16\pi}\int d^4x\sqrt{-g}(R-2(\nabla\varphi)^2+2\lambda\,\varphi\, {\mathcal R}^2_{\rm GB})\,,
\end{eqnarray}
where $g$ is the determinant of the spacetime metric and ${\mathcal R}^2_{\rm GB}=R^2-4R_{\mu\nu}R^{\mu\nu}+R_{\mu\nu\rho\sigma}R^{\mu\nu\rho\sigma}$ is the Gauss-Bonnet curvature invariant, coupled to the scalar field $\varphi$ via a coupling constant $\lambda$ with dimensions of $[{\rm length}]^2$.
All BHs in this theory are known to possess a non-trivial scalar field configuration, known as scalar hair, with an asymptotic fall-off $\varphi=\varphi_0+Q_{\rm SF}/r+{\mathcal O}(1/r^2)$ . Here, $\varphi_0$ is the scalar field value at spatial infinity\footnote{We set $\varphi_0=0$, appropriate for asymptotically flat spacetimes as considered here; this will not hold for a non-trivial cosmological background.} and $Q_{\rm SF}$ its scalar charge.

The scalar field effects are encoded in dimensionless, body-dependent parameters, known as sensitivities, which describe the adiabatic response of compact object's mass and scalar charge. 
For non-spinning BHs in shift-symmetric EsGB, the sensitivities $\alpha_i$ and its derivatives $\beta_i$ and $\beta_i'$ up to 3PM order are given by~\cite{Julie:2019sab,Julie:2022huo}
\begin{align}
    \alpha_i&=-\frac{\lambda}{\mu_i^2}\,,\nonumber\\
    \beta_i&=\left.\frac{d\alpha_i}{d\varphi}\right|_{\varphi_0}=-\frac{2\lambda^2}{\mu_i^4}\,,\nonumber\\
    \beta_i'&=\left.\frac{d\beta_i}{d\varphi}\right|_{\varphi_0}=-\frac{8\lambda^3}{\mu_i^6}\,,
    \label{eq:couplings}
\end{align}
with $\{\mu_i\}_{i=1,2}$ being the irreducible BH masses. 

The most stringent observational bound on EsGB theory comes from comparing the black hole-neutron star GW signal GW230529 to PN results and give a constraint of $\sqrt{\lambda}\lesssim 0.75$ km~\cite{Sanger:2024axs}. For comparison, assuming an equal-mass system with the component BH masses being $3.6M_{\odot}$ (the smallest BH observed so far~\cite{Song:2024tqr}),
the highest coupling value considered here is $\sqrt{\lambda}\sim 1.9$ km, which is above the observational constraints but remains within the same order of magnitude (see \cite{Corman:2025wun} for further details).

\vskip 5pt

{\textit{\textbf{PM–EOB analytic mapping of {the scattering angle}}---}
Recently, there has been a significant effort in comparing analytical predictions with numerical relativity simulations for black hole scattering configurations~\cite{Hopper:2022rwo,Damour:2022ybd,Rettegno:2023ghr,Swain:2024ngs,Buonanno:2024vkx,Long:2025nmj}. Following Ref.~\cite{Damour:2022ybd}, we analytically describe the scattering of BHs in EsGB theories by transcribing the PM-expanded scattering angle $\chi(\gamma, j)$ (where $\gamma$ is the Lorentz factor and $j= J/(Gm_1m_2)$
is the dimensionless orbital angular momentum, with $\{{m}_i\}_{i=1,2}$ the component mass of BHs) into a corresponding energy-dependent EOB radial potential $w(\gamma, \bar{r})$ (where $\bar{r} = \bar{R}/(GM)$ is the dimensionless isotropic EOB radial coordinate, with $M=m_1+m_2$). The corrections due to EsGB enter perturbatively with respect to GR, the scattering angle can therefore be decomposed as
\begin{align}
\chi^{\rm BH}_{\rm EsGB} = \chi^{\rm BH}_{\rm \small{GR}} + \chi^{\rm BH}_{\rm \small{mod}}~,     
\end{align}
leading to a corresponding decomposition of the radial potential
\begin{align}
   w^{\rm BH}_{\rm EsGB} = w^{\rm BH}_{\rm \small{GR}} + w^{\rm BH}_{\rm \small{mod}}~. 
\end{align}
The PM expansion of $\chi$ uniquely determines the PM expansion of $w$ via the EOB-derived map \cite{Damour:2017zjx}
\begin{align}
  \label{eq:chirelation}
  \pi+\chi(\gamma,j) = 2 j \int_0^{\bar{u}_{\rm max}(\gamma, j)}
  \!\!\!\!
  \frac{d\bar{u}}{\sqrt{p_{\infty}^2+w(\bar{u},\gamma)-j^2 \bar{u}^2}}~,
\end{align}
where $\bar{u}=1/\bar{r}$\,, $\bar{u}_{\rm max} \equiv 1/\bar{r}_{\rm
min}$ and $p_{\infty}^2=\gamma^2-1$. Here, $\bar{r}_{\rm min}$ is the radial turning point, defined as the largest root of the (total) radial potential, $p_{\bar{r}}^2 = {p_{\infty}^2+w(\bar{r},\gamma)-j^2/ \bar{r}^2}$. Upon PM expanding both sides of Eq.~\ref{eq:chirelation}, one performs the subsequent (partie finie) integrals to determine the coefficients of $w$. The EOB-resummed scattering angle, $\chi^{w-\mathrm{EOB}}_{n\mathrm{PM}}$, is then obtained by evaluating the integral in Eq.~\ref{eq:chirelation}, with $w$ specified up to $n$PM order. In GR, for non-spinning BHs, the 4PM knowledge of $\chi^{\rm BH}_{\rm GR}$~\cite{Kalin:2020fhe, Bern:2019nnu, Bern:2021yeh, Dlapa:2021vgp, Bjerrum-Bohr:2021din} 
determines completely $w^{\rm BH}_{\rm GR}$ up to 4PM order
\cite{Damour:2022ybd}, complemented by a numerically
fitted 5PM contribution, $w_5(\gamma)^{\rm BH, NR}_{\rm GR}$ \cite{Rettegno:2023ghr, Swain:2024ngs}. 

The state-of-the-art of the scattering angle $\chi^{\rm BH}_{\rm \small{EsGB}}$ in EsGB and massless scalar-tensor theories is up to 3PM \cite{Bernard:2026old,Almeida:2026abr} and 3PN order \cite{Jain:2023vlf} for the conservative dynamics. Using the relation between $w$ and $\chi$, we derive the modification in the $w$-potential due to EsGB. The explicit expressions of 1PM and 2PM corrections are
\begin{align}
    w^{\rm BH}_{\small{\rm 1PM, mod}} = &~ 2 \alpha_1 \alpha_2~, \nonumber\\
    w^{\rm BH}_{\small{\rm 2PM, mod}} = &~ \frac{1}{2 h} \left\{8 \alpha_1\alpha_2 + 4 (1+2\alpha_1 \alpha_2) \left(\frac{m_2}{M} \beta_1 + \frac{m_1}{M} \beta_2\right)
    \right. \nonumber\\
    &\left. + \frac{m_2}{M}\alpha_2^2 \left(\gamma^2-1 + 4 \alpha_1^2(1+\beta_1)\right)     \right. \nonumber\\
    &\left. + \frac{m_1}{M}\alpha_1^2 \left(\gamma^2-1 + 4 \alpha_2^2(1+\beta_2)\right) \right\}~,
\end{align}
where $h = \sqrt{1+2\nu(\gamma-1)}$ with the symmetric mass ratio $\nu=m_1m_2/M^2$. The full analytical expression of the conservative part of $w^{\rm BH, cons}_{\small{\rm 3PM, mod}}$ is given in Appendix~\ref{app:3pm}. The incorporation of these EsGB corrections modifies non-trivially the scattering dynamics. In particular, additional contributions to $w$ alter the root structure of $p_{\bar{r}}^2$, leading to changes in the critical angular momentum with respect to GR.

The radiation effects in EsGB theories also include dipolar contributions, which can be significant and should, in principle, be taken into account while computing the analytical estimates of the scattering angle. However, for the equal-mass cases considered here, it is sufficient to include only the GR radiation effect contributions. This is further supported by the excellent agreement between analytical and numerical scattering shown in Fig.~\ref{fig:compareNRvsAR}.

\vskip 5pt
{\textit{\textbf{Numerical simulations}}---}
We use the publicly available code, \texttt{GRFolres} \cite{AresteSalo:2023hcp}, an extension of \texttt{GRChombo} \cite{Clough:2015sqa, Andrade:2021rbd} to four-derivative scalar-tensor (4$\partial$ST) theories of gravity. We evolve the EsGB equations using mCCZ4 formalism \cite{AresteSalo:2022hua,AresteSalo:2023mmd}, which utilizes the modified harmonic gauge (MHG) introduced in \cite{Kovacs:2020pns, Kovacs:2020ywu} to maintain well-posedness in the weakly coupled regime. We choose the domain size $L = 1024M$, with $N = 160$ grid points along every Cartesian coordinate on the coarsest refinement level. Adaptive mesh refinement is based on the hybrid tagging criterion proposed in \cite{Radia:2021smk}, where the near-zone levels are updated based on a pre-determined radius from each BH puncture, while the wave-zone levels are dependent on the second derivative of the conformal factor. The radius of our finest levels is set to $2M$, as we find that this improves the accuracy of our results~\footnote{For simulations that explore stronger gravitational interactions than those considered here, larger radii are required.}. The computational grid possesses 10 levels of refinement, leading to a resolution of $\Delta x = M/80$ around each black hole on the finest level. We choose the constraint damping parameters $\kappa_1 = 0.6/M$, $\kappa_2 = -0.15$, and $\kappa_3 = 1.0/M$, together with the modified CCZ4 parameters $a_0 = 0.1$ and $b_0 = 0.2$, following \cite{Capuano:2026lhs}.

We construct initial data as follows. Since quasi-stationary initial data for binary black holes in EsGB are not yet available (see Refs.~\cite{Brady:2023dgu,Nee:2024bur} for recent progress), we solve the GR constraint equations using the \texttt{TwoPunctures} code to generate Bowen-York initial data for non-spinning, equal-mass black holes~\cite{Damour:2014afa}. The scalar field is initialized trivially, $\varphi=\partial_t\varphi=0$, which leads to an initial unphysical transient before settling to an EsGB configuration. The momenta of black holes are
\begin{align}
    \vec{P} = \pm P_\mathrm{ADM} \left( - \sqrt{1 - \frac{b_\mathrm{NR}}{D}}, \frac{b_\mathrm{NR}}{D}, 0 \right)~, 
\end{align}
where $b_\mathrm{NR}$ is related to the ADM angular momentum via $J_\mathrm{ADM} = P_\mathrm{ADM} b_\mathrm{NR}$. We choose an initial separation $D = 100M$ and $P_\mathrm{ADM} = 0.1145639M$.

We perform a suite of five simulations exploring the parameter space $\mathcal{P} \in \{\lambda/M^2, b_\mathrm{NR}/M\}$; details are given in Table~\ref{tab:initData}. We extract the scattering angles following the polynomial fitting procedure laid out in \cite{Damour:2014afa}, with the fitting windows for the ingoing and outgoing trajectories being  $r \in [20,90]M$ and  $r \in [30,100]M$, respectively. 
We estimate the errors in the scattering angle due to polynomial-fitting of the trajectories (following Ref.~\cite{Damour:2014afa}) and finite-resolution effects. As shown in Appendix~\ref{app:conv_test}, the resolution errors are subdominant, with differences in the scattering angle on the order of $10^{-1}$ degrees. 
Therefore, we only report errors due to polynomial-fitting in Table~\ref{tab:initData}.

As a validation of our computational setup, we perform a dedicated simulation with parameters $\{0, 9.7\}$ to recover the GR limit. 
We find excellent agreement with the corresponding $\texttt{Einstein Toolkit}$ simulation of \cite{Swain:2024ngs}, where $\chi_\mathrm{GR}^\mathrm{ETK} - \chi_\mathrm{GR}^\mathrm{GRFolres} = 0.569^\circ$, which is within their corresponding error bars. 

We also assess the impact of our initial data on the ADM quantities. As a consequence of initializing the scalar field as $\varphi=\partial_t\varphi = 0$, the scalar field undergoes a period of excitation at the start of a simulation until a quasi-stationary state is obtained. For bound systems, this transient results in perturbed BH trajectories at early times and, consequently, increased eccentricity~\cite{AresteSalo:2025sxc, Corman:2025wun}. However, for the unbound configurations considered here, in particular for simulations $\{0, 9.7\}$ and $\{0.0325, 9.7\}$, we find minimal differences in $E_\mathrm{ADM}$ and $J_\mathrm{ADM}$ between GR and EsGB, after extraction on fixed coordinate spheres and extrapolation to spatial infinity. This is further supported by the negligible impact on the ingoing trajectories, with the system settling to a quasi-stationary state at a sufficiently early time. 
Additionally, due to the assumption of conformal flatness in constructing Bowen-York data, an initial burst of so-called `junk' radiation is emitted. As in \cite{Damour:2014afa}, we find that this has a negligible impact on our initial ADM quantities. We therefore need not apply any corrections to the ADM quantities due to our imperfect initial data.

\begin{table}[htbp]
    \setlength{\tabcolsep}{0.09cm}
    \begin{tabular}{l | c| c| c| c| c}
    \multicolumn{1}{c|}{\fontsize{7.5}{11}\selectfont $b_\mathrm{NR}/M$} & \multicolumn{1}{c|}{\fontsize{7.5}{11}\selectfont $E_\mathrm{ADM}/M$} & \multicolumn{1}{c|}{\fontsize{7.5}{11}\selectfont $J_\mathrm{ADM}/M^2$} & \multicolumn{1}{c|}{\fontsize{7.5}{11}\selectfont $\lambda/M^2$}  & \multicolumn{1}{c|}{\fontsize{7.5}{11}\selectfont $\mu_i/M$} & \multicolumn{1}{c}{\fontsize{7.5}{11}\selectfont $\chi_{\rm NR}(\Delta{\chi}_{\rm NR})$ [deg]} \\
    \hline
\multicolumn{1}{c|}{\vspace{-8.5pt}} & 
\multicolumn{1}{c|}{} & 
\multicolumn{1}{c|}{} & 
\multicolumn{1}{c|}{} & 
\multicolumn{1}{c|}{} & 
\multicolumn{1}{c}{} \\
    \hline
    9.7 & 1.02264 & 1.11128 & 0.0000 & 0.5000 & $273.798\, (0.893)$ \\
    9.7 & 1.02264 & 1.11128 & 0.0175 & 0.5005 & $275.357\, (1.043)$ \\
    9.7 & 1.02264 & 1.11128 & 0.0250 & 0.5010 & $277.715\, (1.193)$\\
    9.7 & 1.02264 & 1.11128 & 0.0325 & 0.5017 & $280.588\, (1.200)$ \\
    11.0  & 1.02264 & 1.26021 & 0.0250 & 0.5010 & $152.263\, (0.649)$ \\
    \hline    
    \hline
    \end{tabular}
    \caption{Summary of the initial data and scattering angle for all simulations considered in this work. Note that the symmetric errors quoted on the scattering angle are from the extrapolation procedure only, since resolution errors are sub-dominant (see Tab.~\ref{tab:conv_test}). }
    \label{tab:initData}
\end{table}

\vskip 5pt
{\textit{\textbf{Comparison between analytical and numerical results}}---}
The analytical predicted angle, defined by Eq.~\ref{eq:chirelation}, depends on the knowledge
of the following quantities: $E_{\rm ADM}$ (or equivalently $\gamma=2(E_{\rm ADM}/M)^2-1$),
$J_{\rm ADM}$, $w_5^{\rm BH}$, and the
sensitivity parameters $\alpha_i$, $\beta_i$, $\beta'_i$. 
The determination of the initial $E_{\rm ADM}$ and $J_{\rm ADM}$ values from the NR simulations is based on the initial data construction described above (see Table~\ref{tab:initData}).  
We now in turn discuss the derivation of $w_5^{\rm BH}$ and sensitivity parameters $\alpha_i$, $\beta_i$, $\beta'_i$.

We use the results of Ref.~\cite{Rettegno:2023ghr}, evaluated at the same initial energy as in our setup. The corresponding effective 5PM contribution to the BH potential $w^{\rm BH}(\bar r,\gamma)$ at this energy is
\begin{align}
w_5^{\rm BH} = -1.00 \pm 0.17~.
\end{align}

Next, we determine the irreducible masses $\mu_i$ from the simulations via the Wald entropy of the BHs, $\mu_i = \sqrt{{\mathcal S}_{\rm W}/(4\pi)}$~\cite{Julie:2019sab,Julie:2022huo}, where ${\mathcal S}_{\rm W}=A_{\rm H}/4+8\pi\lambda\,\varphi_{\rm H}$ is determined through the area of the apparent horizon $A_{\rm H}$, as well as the scalar field value $\varphi_{\rm H}$ on it~\footnote{In particular, we use the surface-averaged value of $\varphi_\mathrm{H}$ in this computation.}. As shown in Fig.~\ref{fig:compareirrmasses}, the value of the irreducible masses changes slightly after scalarization but remains constant thereafter. In Table~\ref{tab:initData} we present its value after scalarization for each simulation, which we use 
together with the coupling strengths $\lambda/M^2$
and Eq.~\ref{eq:couplings} to compute the corresponding sensitivities.

We then compare our analytical prediction with the full set of five NR scattering data {$\chi_{\rm NR}$} (see Table \ref{tab:initData})
composed of three types of Gauss-Bonnet coupling strength. The analytical prediction ($\chi_{\rm EsGB}^{\rm BH}$, $w_{\rm EsGB}^{\rm BH}$) is obtained by adding the EsGB corrections ($\chi_{\rm mod}^{\rm BH}$, $w_{\rm mod}^{\rm BH}$) order by order to the 5PM GR results ($\chi_{\rm GR}^{\rm BH}$, $w_{\rm GR}^{\rm BH}$).
Fig.~\ref{fig:compareNRvsAR} presents a comparison of our analytical predictions (solid lines) with the corresponding NR data (circles) for $b_{\rm NR}=9.7M$ and different coupling strengths. The analytical predictions are presented in their resummed EOB potential $w$ representation. 
For the impact parameter $b_{\rm NR}=9.7M$, where the scattering angle is large ($\chi > 200^\circ$), the agreement is excellent using the analytic description based on the $w$-resummation for all coupling strengths.
\begin{figure}[t]
  \centering
  \includegraphics[width=0.95\linewidth]{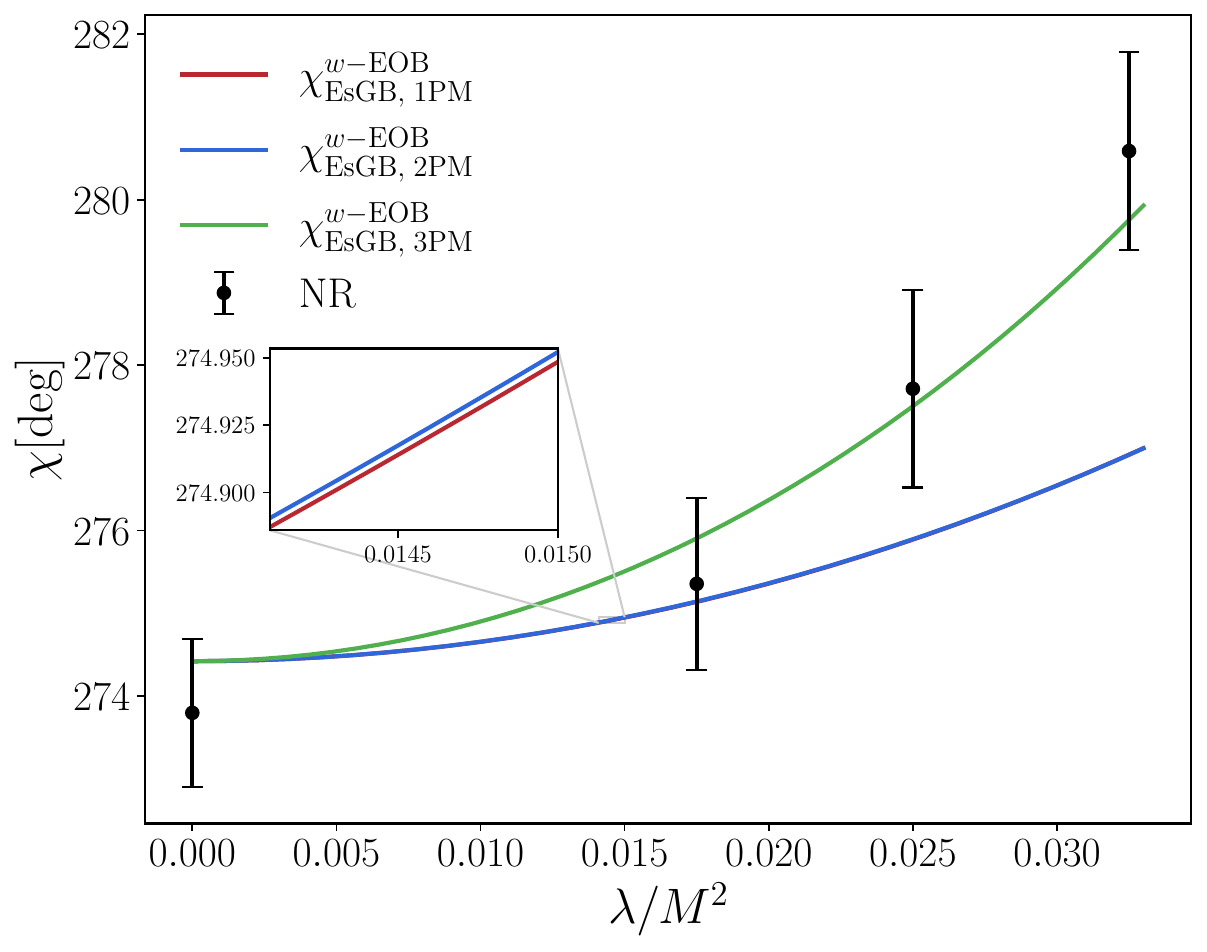}
  \caption{Scattering angle comparison for impact parameter $b_{\rm NR}=9.7M$ between the NR data (filled circles) and the
  EOB-resummed, PM based analytical prediction (solid lines) for equal-mass,
  non-spinning black hole configurations at various Gauss-Bonnet coupling strengths in EsGB.
  }
  \label{fig:compareNRvsAR}
\end{figure}
\begin{figure}[t]
  \centering
  \includegraphics[width=\linewidth]{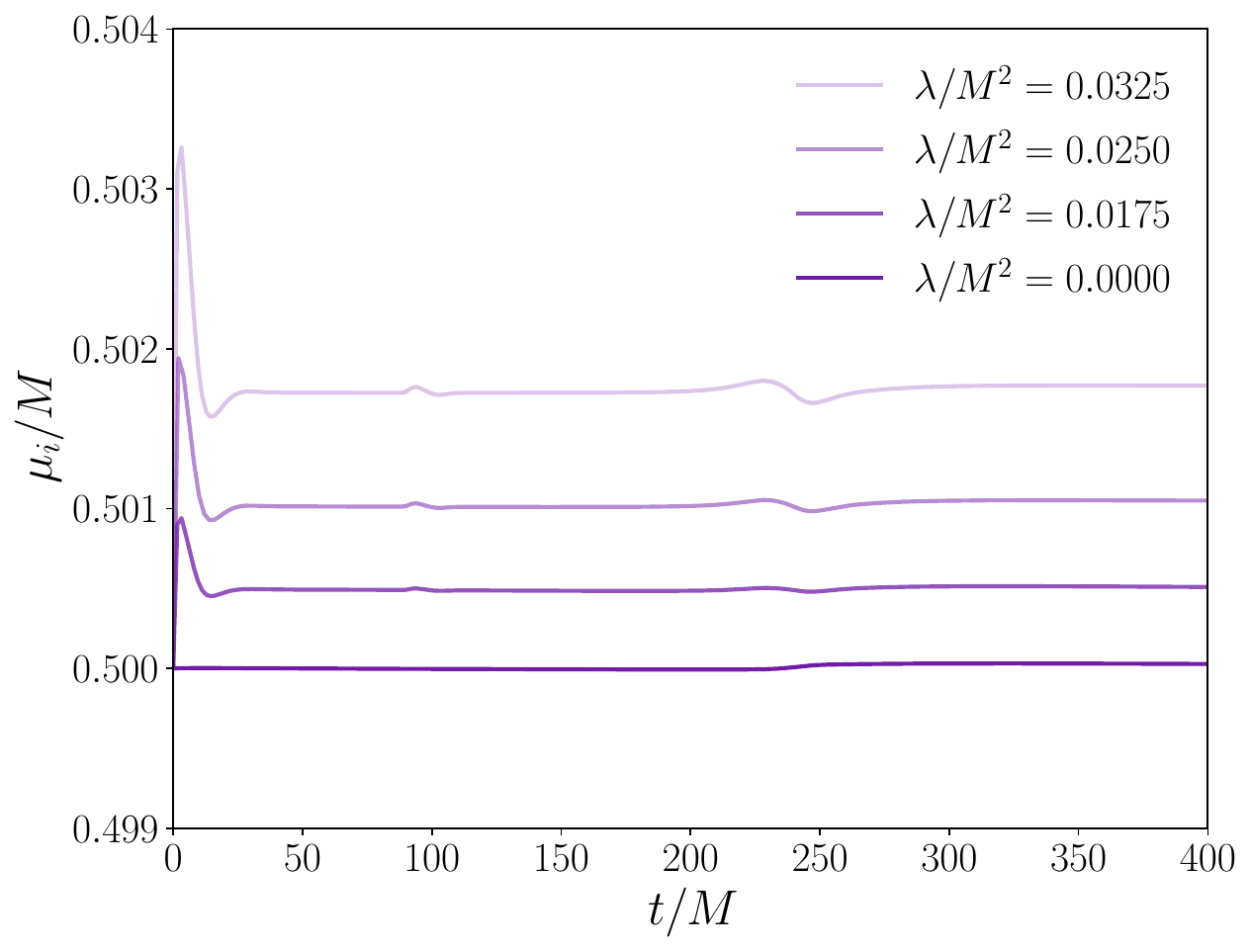}
  \caption{Evolution of the irreducible mass, $\mu_i$, for all simulations in the $\{\lambda/M^2, 9.7\}$ series. For a non-zero EsGB coupling there is an artificial jump in $\mu_i$ due to the imperfect initial data, before settling down to a stationary value.
  }
  \label{fig:compareirrmasses}
\end{figure}
Additionally, we find that at a larger impact parameter, $b_{\rm NR}=11.0M$, for $\lambda=0.0250M^2$ the results (both analytical and NR) are closer to GR, indicating that EsGB effects are suppressed in this regime. Further details about the NR/analytical comparison are given in
Table~\ref{tab:ARData}, which explores the effect of successive PM orders in the $w$-resummation form. 

We also define the quantify $\delta_{\chi} \defeq (\chi_{\rm analytical} - \chi_{\rm NR})/\Delta
\chi_{\rm NR}$ to measure the deviation of analytical results from the NR results in units of the NR error estimate $\Delta\chi_{\rm NR}$. The results for various $w$-resummed analytical estimates are displayed in Table~\ref{tab:ARData}. {In the strong-coupling regime, the 3PM result shows a significant improvement over the 2PM prediction. Moreover, $\delta_{\chi}^{\rm 3PM}$ lies between $[-1,1]$ for all configurations considered, indicating an excellent agreement between analytical and numerical results.} 
 
\begingroup
\begin{table*}[ht]
\begin{center}
\caption
{Summary of the analytical scattering angle for
various EOB predictions ($\chi^{w-\rm EOB}_{\rm \scriptsize{EsGB, nPM}}$) with impact parameter $b_{\rm NR}$ and coupling strength $\lambda$. Here,
$\delta_{\chi} = (\chi_{\rm analytical} - \chi_{\rm NR})/\Delta
\chi_{\rm NR}$ measures the deviation of analytic values
from NR values in units of the NR error estimate.} 
    \setlength{\tabcolsep}{12.5pt}
    \begin{tabular}{l |c |c |c |c |c |c |c  
    }
\multicolumn{1}{c|}{$b_\mathrm{NR}/M$} & \multicolumn{1}{c|}{$\lambda/M^2$} &  
\multicolumn{1}{c|}{$\chi^{ \small{w{\scriptsize -}\rm{EOB}}}_{\rm \scriptsize{EsGB,1PM}}$ }& \multicolumn{1}{c|}{$\delta_{\chi}^{\scriptsize{\rm{1PM}}}$} &  \multicolumn{1}{c|}{$\chi^{ \small{w{\scriptsize -}\rm{EOB}}}_{\rm \scriptsize{EsGB,2PM}}$ } & \multicolumn{1}{c|}{$\delta_{\chi}^{\scriptsize{\rm{2PM}}}$} & \multicolumn{1}{c|}{$\chi^{ \small{w{\scriptsize -}\rm{EOB}}}_{\rm \scriptsize{EsGB,3PM}}$ } & \multicolumn{1}{c}{$\delta_{\chi}^{\rm{3PM}}$}\\
\hline
\multicolumn{1}{c|}{\vspace{-8.5pt}} & 
\multicolumn{1}{c|}{} & 
\multicolumn{1}{c|}{} & 
\multicolumn{1}{c|}{} & 
\multicolumn{1}{c|}{} & 
\multicolumn{1}{c|}{} & 
\multicolumn{1}{c|}{} & 
\multicolumn{1}{c}{} \\
\hline
9.7 & 0.0000 & $274.42^\circ$ &  0.70 &$274.42^\circ$ &  0.70 &$274.42^\circ$ & 0.70  \\
9.7 & 0.0175 & $275.14^\circ$ & -0.21 &$275.15^\circ$ & -0.20 &$275.91^\circ$ & 0.53  \\
9.7 & 0.0250 & $275.89^\circ$ & -1.53 &$275.90^\circ$ & -1.52 &$277.50^\circ$ & -0.18 \\
9.7 & 0.0325 & $276.90^\circ$ & -3.07 &$276.90^\circ$ & -3.07 &$279.72^\circ$ & -0.72 \\
11  & 0.0250 & $152.22^\circ$ & -0.07 &$152.22^\circ$ & -0.06 &$152.28^\circ$ & 0.02 \\
\hline
\hline
\end{tabular}
\label{tab:ARData}
\end{center}
\end{table*}
\endgroup

\vskip 5pt
{\textit{\textbf{Conclusions}}---}
We have performed the first numerical simulations for hyperbolic encounters of 
non-spinning, equal-mass BH binary encounters in EsGB theory 
with numerical computation of scattering angles between $\sim 150\degree$\, and $\sim280\degree$. As a validation of the NR code \texttt{GRFolres}, we also performed a simulation in GR and obtained results consistent with those of \texttt{Einstein Toolkit} \cite{Damour:2014sva, Swain:2024ngs}. 
On the analytic side, our key results include the derivation of the correction 
to the EOB $w$-potential up to 3PM order (conservative) due to scalar-scalar and scalar-gravitational corrections 
in EsGB theories using the PM scattering angle of \cite{Bernard:2026old, Almeida:2026abr}.

The comparison of the numerical/analytical results demonstrate the effect of the scalar field on the dynamics of BHs, and we have found an excellent agreement across all configurations considered, spanning multiple coupling strengths and two distinct impact parameters. In contrast to bound orbits, the absence of fully consistent initial data beyond GR has a negligible effect on both the ingoing trajectories and the ADM quantities. However, the validity of this statement may alter as one explores the parameter space. These results indicate that hyperbolic encounters provide an ideal framework for benchmarking analytic approaches against numerical simulations in higher-derivative theories of gravity. 

As a concluding remark, we note that our results for hyperbolic encounters differ from the recent NR simulations of 
bound systems \cite{Corman:2025wun} for similar coupling constants, 
which were found to evolve more slowly (delayed merger) relative to GR, while analytical models instead predict a faster merger.
This regime dependence {may suggest
that strong scalar coupling effects become important for prolonged strong-field interactions in the bound orbit configurations.
This highlights the need to further explore and understand the transition regime between hyperbolic and bound regimes to clarify this distinct feature.

\vskip 5pt
\noindent
\textbf{\textit{Acknowledgements.}}—
The authors thank Katy Clough, Thibault Damour, Geraint Pratten and Miren Radia for useful discussions during the preparation of this manuscript. 
We thank the GRTL collaboration (\url{www.grtlcollaboration.org}) for their support and code development work. 
S.~S. acknowledges support from a Royal Society University Research Fellowship URF{\textbackslash}R1{\textbackslash}221500 and RF{\textbackslash}ERE{\textbackslash}221015. T.~J. is supported by a LabEx Junior Research
Chair Fellowship at ENS, Paris. L.~A.~S. is partly funded by
Interuniversitaire Bijzonder
Onderzoeksfonds (IBOF)/21/084.
This work was granted access to the HPC resources of MesoPSL financed
by the Region \^Ile de France and the project Equip@Meso (reference
ANR-10-EQPX-29-01) of the programme Investissements d’Avenir supervised
by the Agence Nationale pour la Recherche. This work utilized the resources of the Sulis Tier 2 HPC platform hosted by the Scientific Computing Research Technology Platform at the University of Warwick. Sulis is funded by EPSRC Grant EP/T022108/1 and the HPC Midlands+ consortium, and the University of Birmingham’s BlueBEAR HPC, which provides a High Performance Computing service to the University’s research community. This work also used the DiRAC@Durham facility managed by the Institute for Computational Cosmology on behalf of the STFC DiRAC HPC Facility (\url{www.dirac.ac.uk}). The equipment was funded by BEIS capital funding via STFC capital grants ST/P002293/1, ST/R002371/1 and ST/S002502/1, Durham University and STFC operations grant ST/R000832/1. DiRAC is part of the National e-Infrastructure.

\bibliography{prlref}

\clearpage
\newpage

\section*{Appendices}
\subsection{Correction to the 3PM effective-one-body potential $w$}\label{app:3pm}
The Einstein-scalar-Gauss-Bonnet (EsGB) correction to the conservative sector of the 3PM effective-one-body potential $w$ using the 3PM scattering angle of \cite{Bernard:2026old, Almeida:2026abr} is given by 
\begin{align}
    w_{\rm 3PM, mod}^{\rm BH, cons} =~& \frac{ C^{\rm mod}_{\rm a} }{2h(\gamma^2-1)} +\frac{ C^{\rm mod}_{\rm b} }{6h^2(\gamma^2-1)^2}   
    \nonumber \\
    &+ \nu \frac{\arcosh({\gamma})}{h^2 p_{\infty}} C^{\rm mod}_{\rm c}~,
\end{align}
where $p_{\infty} = \sqrt{\gamma^2-1}$, $h = \sqrt{1+2\nu(\gamma-1)}$ with Lorentz factor $\gamma$ and symmetric mass ratio $\nu = m_1 m_2/(m_1+m_2)^2$. The coefficients $C^{\rm mod}_{i}$ ($i=\mathrm{a,b,c}$) encode the corrections due to EsGB and are given explicitly by
\begin{widetext}
\begin{align}
        C^{\rm mod}_{\rm a} =~& \aA \aB (11-31\gamma^2 - 8 \aA\aB) + \frac{m_2}{M} \bigg(1-2\gamma^2-\aA\aB\bigg) \bigg[4 \bA (1+2\aA \aB) + \aB^2 (\gamma^2-1+4\aA^2(1+\bA))\bigg]
        \nonumber \\
        &+ \frac{m_1}{M} \bigg(1-2\gamma^2-\aA\aB\bigg) \bigg[4 \bB (1+2\aA \aB) + \aA^2 (\gamma^2-1+4\aB^2(1+\bB))\bigg]~,\nonumber\\
        C^{\rm mod}_{\rm b} =~&\frac{m_2}{ \dB M} \left\{\aA \aB^3 \left[2 \aA^2 \dB \left(\gamma^2\left(\gamma ^2-2\right) (\epsilon_1-8)+\epsilon_1-11\right)+ \left(12 \bA +3 \gamma_{12}^2\right) \left(2 \gamma ^4-5 \gamma ^2+3\right)-8 \left(\gamma ^2-1\right)^2 \dB\right] \right.
        \nonumber\\
        &\left.-\aB^2 \left[\left(12 \bA + 3 \gamma_{12}^2\right) \left(4 \gamma ^4-7 \gamma ^2+3\right)+8 \left(\gamma ^2-1\right)^2 \left(2 \gamma ^2-1\right) \dB -6 \aA^2 \dB \left( \epsilon_1(\gamma ^4 -2 \gamma ^2 +1)- (2\gamma^2+1)\right) \right]\right.
        \nonumber\\
        &\left.+2 \aA \aB \dB \left(20 \gamma ^4+3 \left(\gamma ^2-1\right)^2 \epsilon_1-46 \gamma ^2+17\right)+2 \left(\gamma ^2-1\right)^2 \dB \epsilon_1\right\}   
        &\nonumber \\
        &+\frac{m_1}{ \dA M}\left\{\aA^3 \aB \left[2 \aB^2 \dA \left(\gamma^2\left(\gamma ^2-2\right) (\epsilon_2-8)+\epsilon_2-11\right)+ \left(12\bB+3\gamma_{12}^2 \right)  \left(2 \gamma ^4-5 \gamma ^2+3\right)-8 \left(\gamma ^2-1\right)^2 \dA\right]\right.
        \nonumber\\
        &\left.-\aA^2 \left[-6 \aB^2 \dA \left(\gamma ^4 \epsilon_2-2 \gamma ^2 (\epsilon_2+1)+\epsilon_2-1\right)+ \left(12 \bB +3 \gamma_{12}^2\right)\left(4 \gamma ^4-7 \gamma ^2+3\right)+8 \left(\gamma ^2-1\right)^2 \left(2 \gamma ^2-1\right) \dA\right]\right.
        \nonumber\\
        &\left. +2 \aA \aB \dA \left(20 \gamma ^4+3 \left(\gamma ^2-1\right)^2 \epsilon_2-46 \gamma ^2+17\right)+2 \left(\gamma ^2-1\right)^2 \dA \epsilon_2 \right\} \nonumber\\
        &-\frac{1}{ \dA \dB} \left\{\aA^3 \aB \dB \nu\left[2 \left(\gamma ^2-1\right)^2 \dA   \left(\aB^2 (\epsilon_1+\epsilon_2-16)+6 \gamma -4\right)-6 \aB^2 \dA+3 (\gamma -1)^2 (\gamma +1) (2 \gamma +3)   \left(4 \bB+\gamma_{12}^2\right)\right] \right.
        \nonumber\\
        &\left.+\aA \aB \dA (\gamma^2 -1) (\gamma -1)\nu   \left[\aB^2 \left(3 (2 \gamma +3) \left(4 \bA+\gamma_{12}^2\right)+4 \left(3 \gamma ^2+\gamma -2\right) \dB\right)+2 \dB ( 52 + 3 (\epsilon_1+ \epsilon_2) (\gamma+1) +28 (1-3 \gamma)) \right] \right.
        \nonumber \\
        &\left. -18 \left(1-2 \gamma ^2\right)^2 \aA\aB\dA \dB +\aA^2 \dB \left[6 \aB^2 \dA \left(3-6 \gamma ^2+(\gamma +1) (\gamma -1)^2  ((\gamma +1)(\epsilon_1+ \epsilon_2)-8)\nu\right)
        \right.\right.
        \nonumber\\
        &\left.\left.+(\gamma -1)^2 (\gamma +1) \left(3 (2 (\gamma -1) \gamma -3) \left(4 \bB+\gamma_{12}^2\right)+4 (\gamma +1) \left(2 (\gamma -2) \gamma ^2+\gamma +2\right) \dA\right)\nu\right]\right.
        \nonumber \\
        &\left.+(\gamma -1)^2 (\gamma +1) \dA \nu  \left[\aB^2 \left(3 (2 (\gamma -1) \gamma -3) \left(4 \bA+\gamma_{12}^2\right)+4 (\gamma +1) \left(2 (\gamma -2) \gamma ^2+\gamma +2\right) \dB\right)+2 (\gamma +1) \dB (\epsilon_1+\epsilon_2\right]\right\}~,\nonumber\\
        C^{\rm mod}_{\rm c} =~& \frac{\aA \aB}{\dA \dB} \Big[ (4\bA+\gamma_{\rm 12}^2)(4\bB+\gamma_{\rm 12}^2) -8\dA \dB (4\gamma^2-4+\aA \aB) \Big]~.
\end{align}
\end{widetext}
Here we have used the following combinations of the sensitivities $\alpha_i\,,\beta_i\,,\beta'_i$;
\begin{align}
    \delta_1   &=  \frac{\alpha_1^2}{(1+\alpha_1\alpha_2)^2}~,& \, \delta_2= \frac{\alpha_2^2}{(1+\alpha_1\alpha_2)^2}~, \nonumber\\
    \epsilon_1 &= \frac{\beta'_1 \alpha_2^3}{(1+\alpha_1\alpha_2)^3}~,& \, \epsilon_2 = \frac{\beta'_2 \alpha_1^3}{(1+\alpha_1\alpha_2)^3}~,\nonumber\\
    \gamma_{12}&= -2\frac{\alpha_1\alpha_2}{(1+\alpha_1\alpha_2)}~.&
\end{align}

\newpage

\subsection{Convergence testing} \label{app:conv_test}

To analyze the convergence of our simulations, we compare the scattering angle for the $\{0.0325, 9.7\}$ configuration at the following resolutions: $\Delta x_\mathrm{NR}^\mathrm{LR} = M/64$, $\Delta x_\mathrm{NR}^\mathrm{MR} = M/80$ and $\Delta x_\mathrm{NR}^\mathrm{HR} = M/96$, which is shown in Table~\ref{tab:conv_test}. We find that the error on the scattering angle due to finite resolution is on the order $\sim0.1^\circ$, which is subdominant to the polynomial errors quoted in Table~\ref{tab:initData}. Therefore, we omit this uncertainty from all analyses in this work.

\begin{table}[htbp]
    \setlength{\tabcolsep}{0.15cm}
    \begin{tabular}{c c}
    \multicolumn{1}{c}{$\Delta x_\mathrm{NR} / M$}  & \multicolumn{1}{c}{$\chi_{\rm NR}$ [deg]} \\
    \hline
    \hline
    1/64 &  280.570 \\
    1/80 & 280.588 \\
    1/96 &  280.674 \\
    \hline    
    \hline
    \end{tabular}
    \caption{ Scattering angle for the $\{0.0325, 9.7\}$ configuration at different finest resolutions
    }
    
    \label{tab:conv_test}
\end{table}

We also study the convergence of the phase of the $(2,2)$ mode of the Newman-Penrose $\Psi_4$ scalar, $\Phi_{22}$, using the convergence factor
\begin{align}
    Q_n=\frac{\big(\Delta x^{\rm LR}_{\rm NR}\big)^n-\big(\Delta x^{\rm LR}_{\rm NR}\big)^n}{\big(\Delta x^{\rm MR}_{\rm NR}\big)^n-\big(\Delta x^{\rm HR}_{\rm NR}\big)^n}.
\end{align}
Our analysis shows a convergence between $4^{\rm th}$ and $6^{\rm th}$ order in the strongest field regime, as depicted in Fig.~\ref{fig:convweyl}. This is consistent with the order of the finite difference stencils used ($6^{\rm th}$), combined with a $4^{\rm th}$ order Runge-Kutta method for time discretization and $4^{\rm th}$ order interpolation between the refinement levels. We note that the convergence order before and after the peak of $\Psi_4$ is closer to $2^{\rm nd}$ order. This is because the value of $\Psi_4$ is several orders of magnitude smaller than around the peak value, and is therefore contaminated by numerical noise.

\begin{figure}[H]
  \centering
  \includegraphics[width=\linewidth]{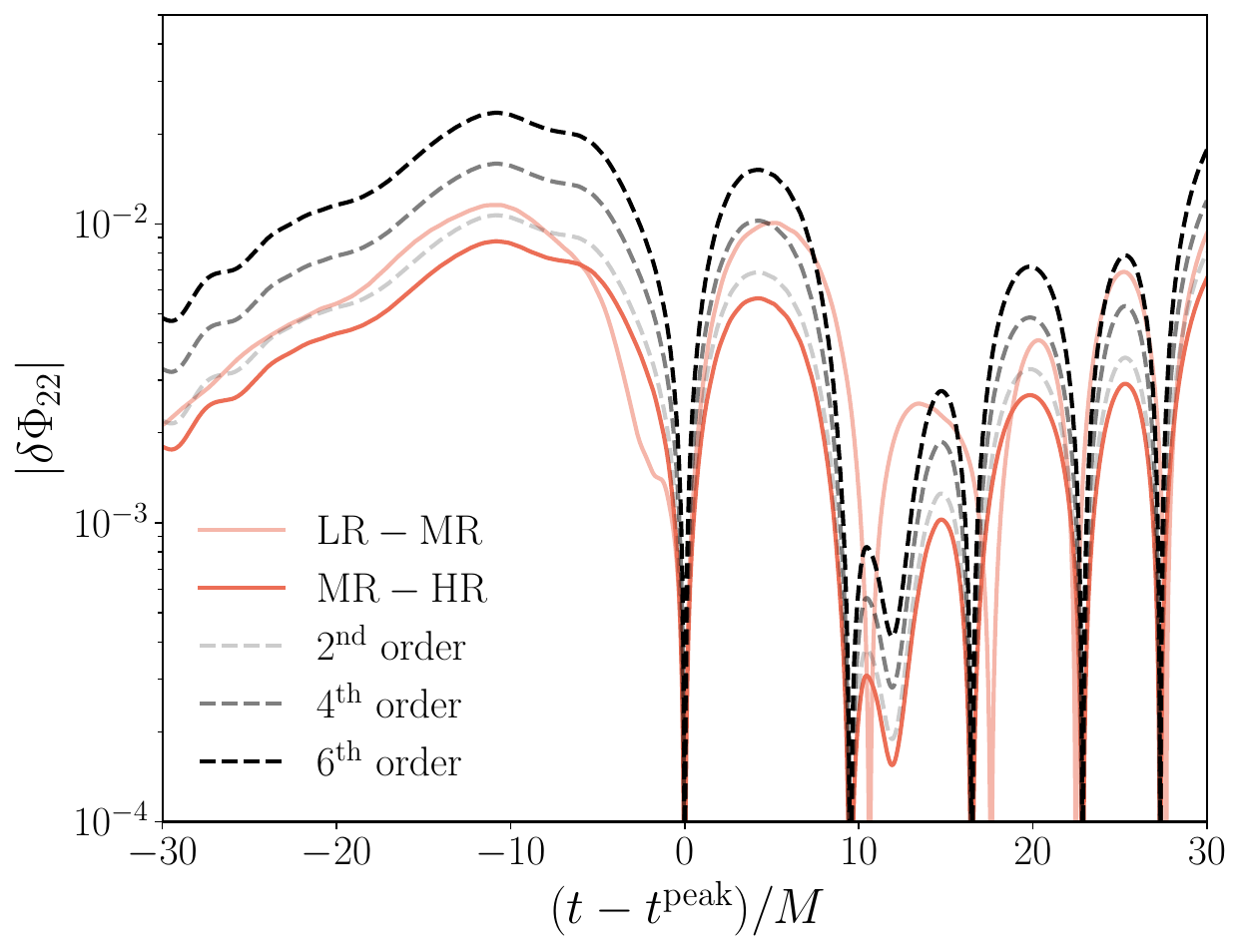}
  \caption{Convergence test for the Newman-Penrose $\Psi_4$ scalar. We show the difference in the values of $\Phi_{22}$ across the three resolutions for the $\{0.0325, 9.7\}$ configuration. The difference between the high
and medium resolution runs has been rescaled by $Q_n$, assuming $2^{\rm nd}$, $4^{\rm th}$ and $6^{\rm th}$ order of convergence.
    } 
  \label{fig:convweyl}
\end{figure}

\end{document}